\documentclass[]{spie}  

\usepackage{booktabs} 
\usepackage{array}
\usepackage{amsmath,amsfonts,amssymb}
\usepackage{float}
\usepackage{graphicx}
\usepackage{subcaption}
\usepackage[colorlinks=true, allcolors=blue]{hyperref}
\title{LLM Gesticulator: Leveraging Large Language Models for Scalable and Controllable Co-Speech Gesture Synthesis}

\author{Haozhou Pang}
\author{Tianwei Ding}
\author{Lanshan He}
\author{Ming Tao}
\author{Lu Zhang}
\author{Qi Gan$^{\ast}$}

\affil{Soul AI, Soulgate Technology Co., Ltd., Shanghai, China}

\authorinfo{${^\ast}$ corresponding author\\Haozhou Pang: E-mail: panghaozhou@soulapp.cn\\  Tianwei Ding: E-mail: dingtianwei@soulapp.cn\\ Lanshan He: E-mail: helanshan@soulapp.cn \\Ming Tao: E-mail: ming@soulapp.cn \\Lu Zhang: E-mail: lu.zhang@soulapp.cn \\Qi Gan: E-mail: ganqi@soulapp.cn}

\begin{document} 
\maketitle

\begin{abstract}
In this work, we present \textit{LLM Gesticulator}, an LLM-based audio-driven co-speech gesture generation framework that synthesizes full-body animations that are rhythmically aligned with the input audio while exhibiting natural movements and editability. Compared to previous work, our model demonstrates substantial scalability. As the size of the backbone LLM model increases, our framework shows proportional improvements in evaluation metrics (a.k.a. scaling law).  Our method also exhibits strong controllability where the content, style of the generated gestures can be controlled by text prompt. To the best of our knowledge, \textit{LLM gesticulator} is the first work that use LLM on the co-speech generation task. Evaluation with existing objective metrics and user studies indicate that our framework outperforms prior works.
\end{abstract}


\begin{figure}[ht]
    \centering
    \includegraphics[width=1\linewidth]{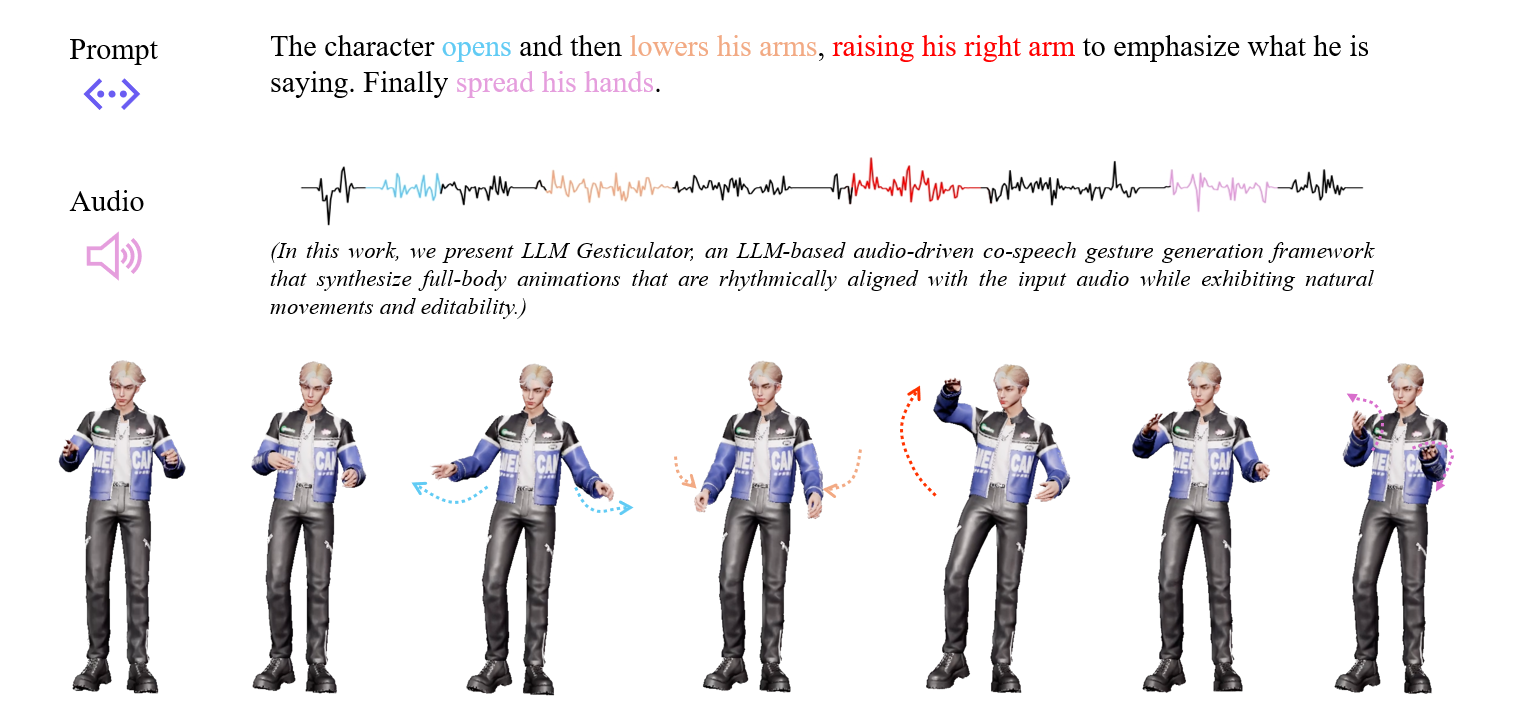}
    \caption{LLM gesticulator synthesizes full-body co-speech gestures according to input audio and text prompt.}
    \label{fig:Banner}
\end{figure}

\section{INTRODUCTION}
\label{sec:intro}  
Gesture, a fundamental aspect of human communication, transcends linguistic barriers and enriches the conveyance of thoughts, emotions, and intentions. It is an intricate movement of the human body that complements speech, providing a nuanced layer of non-verbal interaction \cite{Kendon04}. The significance of gesture in communication is underscored by its role in emphasizing speech content, clarifying complex ideas, and facilitating social cohesion \cite{Macneill92}. The accurate depiction of gestures is paramount for creating immersive and interactive experiences in various sectors, including gaming, film production, robotics, and virtual reality \cite{Bickmore01}.

The quest to generate realistic co-speech gestures has led to the development of rule-based and data-driven methods. Rule-based approaches often rely on predefined rules that govern the synchronization of gestures with speech patterns \cite{Pelachaud95}. Conversely, data-driven methods leverage machine learning to learn the complex mappings from speech to gestures, offering a more organic synthesis process \cite{Bhatta21,Kucherenko20}.

In this paper, we propose a novel framework for generating full-body co-speech gestures that are rhythmically aligned with speech while maintaining scalability and editability. Our approach leverages recent advancements in Large Language Models (LLM) by modeling the gesture generation problem as a sequence-to-sequence translation task. To the best of our knowledge, this is the first attempt of tackling the co-speech generation task using LLMs. Our experiment shows that by proper tokenization of input and delicate design of the LLM training framework, our system can generate high quality gestures that are natural, rhythmic and contextually appropriate. Thanks to the text comprehension ability of pre-trained LLM model and our training framework, our system extends the generation editable capability by allowing user to control motion generation by giving text prompt.

In summary, our main contributions are as follows:
   
\begin{itemize}
  \item We present a framework that is built based on a large language model to generate full body (body + hand) gestures according to input audio and text prompt. Our method outperforms prior works on existing evaluation metrics and user studies. 
  \item Our proposed training scheme supports controllable gesture generation based on text prompts.
  \item We propose a new data augmentation paradigm by utilizing rendering engines and VLLM models. We annotate the motion descriptions of BEAT dataset and we will release it to the community for future research.
  

\end{itemize}
   
\section{Related Work}

\subsection{Co-speech Gesture Generation}
\label{cog}
Synthesizing co-speech gestures is an essential task across various fields, such as games, films, robotics, and virtual reality applications, to create interactive and immersive user experiences. Early methods\cite{beat01,Pelachaud2003,Kopp04,neff08} use rule-based approaches to synthesize gestures, which primarily rely on a carefully hand-crafted heuristic to map input speech to a set of appropriate gesture clips. Wagner et al. \cite{WAGNER2014209} provide a comprehensive summary of rule-based methods. Such methods are highly explainable and controllable, however, the generation diversity and naturalness is limited by the design of rules, and adding more gesture units requires extra manual work, which is infeasible for large-scale systems.

To reduce the manual work in the process of designing rules, and improve the generation naturalness, data-driven methods have drawn much attention recently because of its ability to synthesis gestures from implicit representation of training datasets. Nyatsanga et al.  \cite{Nyatsanga_2023} gives a comprehensive review of data-driven methods. Early data-driven methods try to build a statistical system to predict a gesture label that is used to index pre-recorded gesture clips \cite{kopp06, Levine10}. Compared to rule-based methods, statistical approaches make fewer assumptions about the speech-gesture association, but they still require a well crafted collection of gesture units to generate the final motion. Subsequently, the advanced capabilities of deep neural networks enable the training of models directly using raw gesture data in an end-to-end manner. The development of deep-learning based gesture generation systems is analogous to that of other generation tasks. We briefly introduce some representative works using different network architectures. Deterministic models including MLP \cite{Kucherenko_2020}, CNN \cite{habibie21}, RNN\cite{Bhattacharya21,Hasegawa18,liu22,Yoon_2020,Ao_2022}, and transformers \cite{Bhattacharya_2021} have been studied in previous works. Such deterministic methods tend to produce `mean' pose due to the one-to-many (there could be multiple reasonable gestures for the same input speech) nature of the gesture generation task. To alleviate this problem, generative models have been studied extensively. To name a few, VAEs \cite{li2021audio2gesturesgeneratingdiversegestures,xu2022freeformbodymotiongeneration}, VQ-VAEs \cite{Ao_2022}, flow-based models \cite{alexanderson2020style}, and diffusion-based models \cite{ConvoFusion24} have been studied in previous works. Additionally, some works opt to enhance the semantic correspondence between speech and gestures. Gao et al.  \cite{Gao_2024} use LLM as a classifier to transform the gesture generation into an intention classification problem, the identified gestures are merged with gestures generated from neural network to increase semantic correspondence. Along this line, Zhang et al.  \cite{zhang24} propose a semantic dataset including over 200 types of semantic gestures and fine-tune an LLM to retrieve suitable gesture clips according to the input text. 

With the rapid development of large language models, numerous academic studies have utilized these models to achieve breakthroughs across various fields. To name a few, AnyGPT\cite{zhan2024anygpt} has demonstrated that different modalities including text, audio, and images can be unified using an LLM model to achieve any-to-any multimodal conversation. MotionGPT \cite{jiang2024motiongpt} has shown that relevant tasks can be tackled by an LLM model by treating human motion as a specific language. Our method adopts similar modeling paradigm by using the power of pre-trained LLM models to tackle the co-speech gesture synthesis problem. There are several key differences between \textit{LLM gesticulator} and prior co-speech gesture generation works: (a) we formulate the co-speech gesture generation task as seq2seq translation problem and our system was built based on a pre-trained large language model, we show that such design enables our framework to achieve better performance on evaluation metrics while maintaining scalability. (b) We have explored a data augmentation paradigm in the field of motion generation by annotating the BEAT dataset and use motion content description as text prompt in our training pipeline, which is shown to be effective for a better generation controllability.


\subsection{Dataset}

The volume and quality of datasets is fundamental to any data-driven method. Current public available datasets can be categorized into two groups, pose-estimated from monocular video and motion-captured. The former use 3D pose estimation algorithm to extract motions from videos, Speech2Gesture \cite{ginosar2019gestures} and TED \cite{yoon2018robotslearnsocialskills} estimate the 2D poses first, and then lift to 3D poses and SMPL-X \cite{SMPL-X:2019}. Although pose-estimated methods largely enrich the diversity of motion data, but they are generally limited by the motion quality, and do not contain fingers animations. Motion-captured data usually have better motion quality, Trinity dataset \cite{IVA:2018} propose 4 hours of motion data performed by a male actor along with audio data. The TalkingWithHand \cite{talkwithhand} dataset record conversational motions from two speakers. The ZEGGS dataset \cite{ghorbani2022zeroeggs} propose 2 hours of motion data captured from a single speaker performing with 12 different styles. BEAT dataset \cite{liu2022beat} is by far the largest mocap dataset, it contains 76 hours motion data along with facial blend-shapes and emotions. 

To the best of our knowledge, none of the previous dataset contains audio, motion, and text annotations (describing the content of motion) at the same time. In our work, we render the motion clips in BEAT dataset and use a VLLM model through prompt engineering to describe and summary the detailed motions in each clip (Fig. \ref{fig:vllm}). Thus, another modality  is added to the BEAT dataset. The text description is shown to be useful to extend our framework's controllability such that the style and content of generated gestures can be guided by user's prompt at inference time. We will release the text annotations for BEAT dataset for research purpose.

\subsection{Multi-Modal Data Representation}
Co-speech generation task takes input from multiple modalities, including audio, text, and other speaker-wise information like speaker ids, emotions. The representation and alignment of different modalities have been studied in previous works. For audio input, Mel-spectrogram and MFCC features are commonly used \cite{alexanderson2020style,Kucherenko20,qian2021speech}. As the development of audio compression technology, audio codec is also commonly used in audio-driven tasks for a more abstract and tokenized feature representation \cite{defossez2022highfi}. Text input is usually encoded by pre-trained language models like BERT \cite{devlin2019bertpretrainingdeepbidirectional}. The audio and text features are resampled and aligned into the same length of the motion. 

\subsection{Evaluation of Motion Generation Model}
Evaluating co-speech gestures is challenging due to the stochastic nature of human gesture perception. Most previous works use user studies to evaluate different aspects of motion quality, such as human-likeliness and speech-gesture matching. For quantitative evaluation, serving as supplementary references, previous works \cite{ginosar2019gestures,joo19,Kucherenko_2019} use absolute difference between joint positions, velocity and acceleration between generated motion and ground truth as an indicator of motion quality. However, Kucherenko et al.  \cite{Kucherenko_2024} shows that such scores have no statistically significant correlation with the human-likeness scores from the large user study due to the one-to-many nature of gesture generation task. Fréchet Inception Distance \cite{heusel2018ganstrainedtimescaleupdate} (FID) is a widely used criterion in image generation tasks that measures the difference between the distributions of the dataset and generated samples in the latent space. Yoon et al.  \cite{Yoon_2020} present Fréchet Gesture Distance (FGD) to evaluate the disparity between the latent feature distributions of generated and real gestures, where a lower FGD indicates superior motion quality. 


\section{System Overview}

\label{sec:sections}
\begin{figure}[H]
    \centering
    \includegraphics[width=1\linewidth]{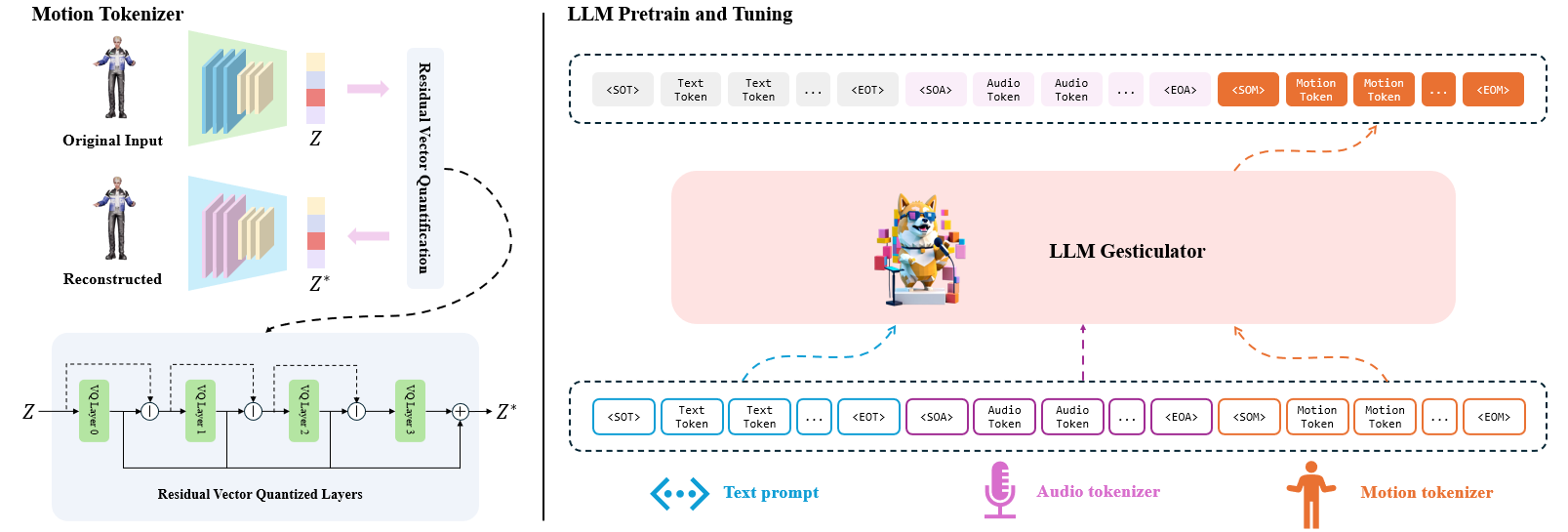}
    \caption{\textbf{System Overview}: Our system consists of two parts. First, we train a Residual VQVAE on motion data to convert it into motion tokens. Then, we use the trained MotionRVQ model to tokenize the motion, and leverage a pre-trained audio tokenizer to convert the audio into tokens. We fine-tune a LLM on these data. The fine-tuned LLM can predict the motion tokens given the audio tokens, and convert them back to a motion sequence through the MotionRVQ Decoder. Text-conditioned generation is achived by adding text tokens in the training pipeline.}
    \label{fig:Sysoverview}
\end{figure}

In this section, we provide a comprehensive overview of how our system processes text prompts and speech audio inputs to predict motion sequences(as shown in Fig. \ref{fig:Sysoverview}). First, we perform motion tokenization, with the detailed methodology outlined in Section \ref{sec:tokenization}. Subsequently, we describe the pretraining of large language models (LLMs) to enable them to comprehend and generate audio data related to motion, as discussed in Section \ref{sec:pretrain}. Finally, we explore the generation of co-speech motion using text prompts, with specific methods and experimental results presented in Section \ref{sec:textprompt}. Our system effectively integrates text guidance with audio-synchronized motion generation, facilitating the production of high-quality motion representations.
 
\subsection{Motion Tokenization}
\label{sec:tokenization}

With the rapid advancement in language model research, LLMs have exhibited exceptional performance. Inspired by the strategies employed in LLMs, we attempt to transform the task of audio-synchronized motion generation into a translation problem solvable by LLMs. Consequently, we adopt the approach of Residual Vector Quantized Variational Autoencoder\cite{lee2022autoregressive} (Residual VQVAE) to construct MotionRVQ for the tokenization of motion data.

For a given motion sequence, we utilize the rotations and offsets of each joint as features, represented as follows:

\begin{equation}
\begin{aligned}
\boldsymbol{m}_{i} &= \{\boldsymbol{R}_0, \boldsymbol{R}_1, \dots, \boldsymbol{R}_N\} \\
\boldsymbol{M} &= \{ \boldsymbol{m}_{0},\boldsymbol{m}_{1},,\dots,\boldsymbol{m}_{L} \}
\end{aligned}
\end{equation}

where $\boldsymbol{R}_i$ denotes the rotation and offset of the $i$-th joint and $\boldsymbol{m}_i$ denotes the motion feature of the $i$-th frame. The MotionRVQ network can be represented as:
\begin{equation}
\boldsymbol{Z} = \boldsymbol{\varPhi}(\boldsymbol{M}),
\end{equation}
where $\boldsymbol{\varPhi}$ is a function representing our MotionRVQ encoder. The tokenization process can be modeled as:

\begin{equation}
    \boldsymbol{Z}^* = \sum_{l=1}^{L} q_l(\boldsymbol{e}_l),
\end{equation}

where $L$ is the number of quantization levels, $q_l(\cdot)$ denotes the quantization function at level $l$, and $\boldsymbol{e}_l$ represents the residual encoding at level $l$. The quantization at each level is defined as:

\begin{equation}
q_l(\boldsymbol{e}_l) = \operatorname{arg\,min}_{\boldsymbol{c} \in \mathcal{C}_l} \left\| \boldsymbol{e}_l - \boldsymbol{c} \right\|^2,
\end{equation}

where $\mathcal{C}_l$ is the codebook at level $l$, and $\boldsymbol{c}$ represents the embeddings in the codebook. The residual encodings are computed recursively as follows:

\begin{equation}
\begin{aligned}
\boldsymbol{e}_1 &= \boldsymbol{\varPhi}(\boldsymbol{M}), \\
\boldsymbol{e}_l &= \boldsymbol{e}_{l-1} - q_{l-1}(\boldsymbol{e}_{l-1}), \quad \text{for } l = 2, \dots, L,
\end{aligned}
\end{equation}

where $\boldsymbol{\varPhi}(\cdot)$ is the encoder function.

Motion tokenizer's architecture comprises convolutional networks and a Transformer Encoder, employing a downsampling rate of 8. Training techniques such as exponential moving average (EMA) and random re-initialization of the inactive codebook entries are used to ensure training stability and enhance quality.

To train our MotionRVQ model effectively, we define a loss function that balances reconstruction accuracy and codebook utilization. The overall loss function $\mathcal{L}$ consists of three components: the reconstruction loss $\mathcal{L}_{\text{rec}}$, the commitment loss $\mathcal{L}_{\text{commit}}$, and the codebook loss $\mathcal{L}_{\text{codebook}}$. The reconstruction loss measures the discrepancy between the original motion sequence $\boldsymbol{M}$ and the reconstructed sequence $\overline{\boldsymbol{M}}$, defined as:

\begin{equation}
    \mathcal{L}_{\text{rec}} = \left\| \boldsymbol{M} - \overline{\boldsymbol{M}} \right\|_2^2,
\end{equation}

where $\overline{\boldsymbol{M}} = \boldsymbol{\varPsi}(\boldsymbol{Z}^*)$, and $\boldsymbol{\varPsi}(\cdot)$ is the decoder function. The commitment loss ensures that the encoder's outputs commit to the codebook entries, encouraging efficient codebook usage. It is defined as:

\begin{equation}
\mathcal{L}_{\text{commit}} = \beta \sum_{l=1}^{L} \left\| \boldsymbol{e}_l - \operatorname{sg}\left[ q_l(\boldsymbol{e}_l) \right] \right\|_2^2,
\end{equation}

where $\beta$ is a hyperparameter controlling the strength of the commitment, and $\operatorname{sg}[\cdot]$ denotes the stop-gradient operation, which prevents gradients from flowing through its argument during backpropagation. The codebook loss updates the codebook entries to match the encoder outputs:

\begin{equation}
\mathcal{L}_{\text{codebook}} = \sum_{l=1}^{L} \left\| \operatorname{sg}\left[ \boldsymbol{e}_l \right] - q_l(\boldsymbol{e}_l) \right\|_2^2.
\end{equation}

The total loss function combines these components:

\begin{equation}
\mathcal{L} = \mathcal{L}_{\text{rec}} + \mathcal{L}_{\text{commit}} + \mathcal{L}_{\text{codebook}}.
\end{equation}

By optimizing this loss function, the model learns to reconstruct motion sequences accurately while effectively utilizing the quantization codebooks. This balance is crucial for generating high-quality motion representations suitable for translation by LLMs.
\subsection{Pretraining the LLM on Motion and Audio Data}
\label{sec:pretrain}
\begin{figure}
    \centering
    \includegraphics[width=0.75 \linewidth]{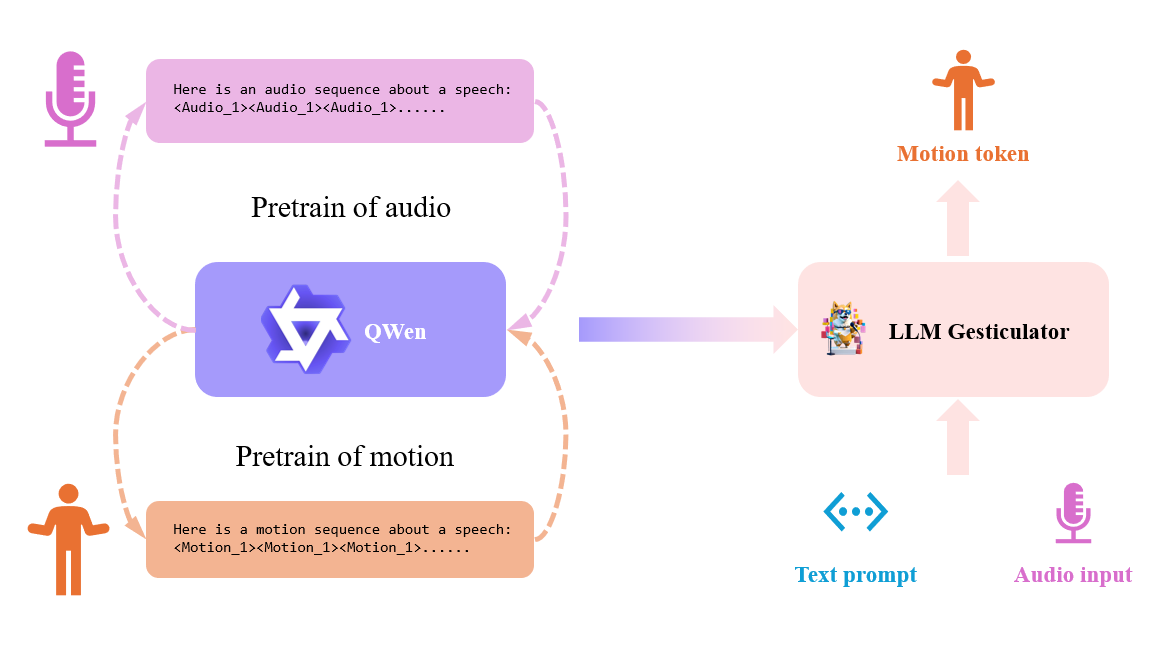}
    \caption{Since the tokens for audio and motion have not been learned by the LLM, it is necessary to pretrain the audio and motion data.}
    \label{fig:pretrainofLLM}
\end{figure}
For the co-speech Motion generation task, alongside the tokenization of motion, it is essential to also tokenize the audio. Recent advancements in audio representation\cite{defossez2022highfi} have led to the development of effective methods for tokenizing speech.

Using the motion tokenizer and audio tokenizer, we convert motion and audio data into discrete token sequences. Let $\hat{\boldsymbol{M}}$ denote the tokenized motion sequence and $\hat{\boldsymbol{A}}$ denote the tokenized audio sequence This transformation allows us to frame audio-driven motion generation as a sequence-to-sequence generation task.

Following prior work~\cite{jiang2024motiongpt}\cite{chen2023executing}, to enhance the LLM's understanding of motion and audio, we first perform motion sequence pre-train and audio sequence pre-train before fine-tuning. For the motion completion task, the model learns to predict the next element in the motion sequence, formalized as estimating the conditional probability $P(\boldsymbol{m}_t \mid \boldsymbol{m}_{1},...,\boldsymbol{m}_{t-1})$, where $\boldsymbol{m}_t$ is the element at time step $t$ and $\boldsymbol{m}_{1},...,\boldsymbol{m}_{t-1}$ is the history of the motion sequence up to time step $t-1$. Similarly, for the audio completion task, the model estimates $P(\boldsymbol{a}_t \mid \boldsymbol{a}_{1},...,\boldsymbol{a}_{t-1})$ to predict the next element in the audio sequence. This process helps the LLM build comprehension of the newly introduced motion and audio tokens.Our experiments (Fig \ref{fig:scale_comparison}.)  demonstrate that this process can enhance the generation quality.

\subsection{Text Prompted Co-Speech Motion Generation using LLM}
\label{sec:textprompt}
\begin{figure}[ht]
    \centering
    \includegraphics[width=1\linewidth]{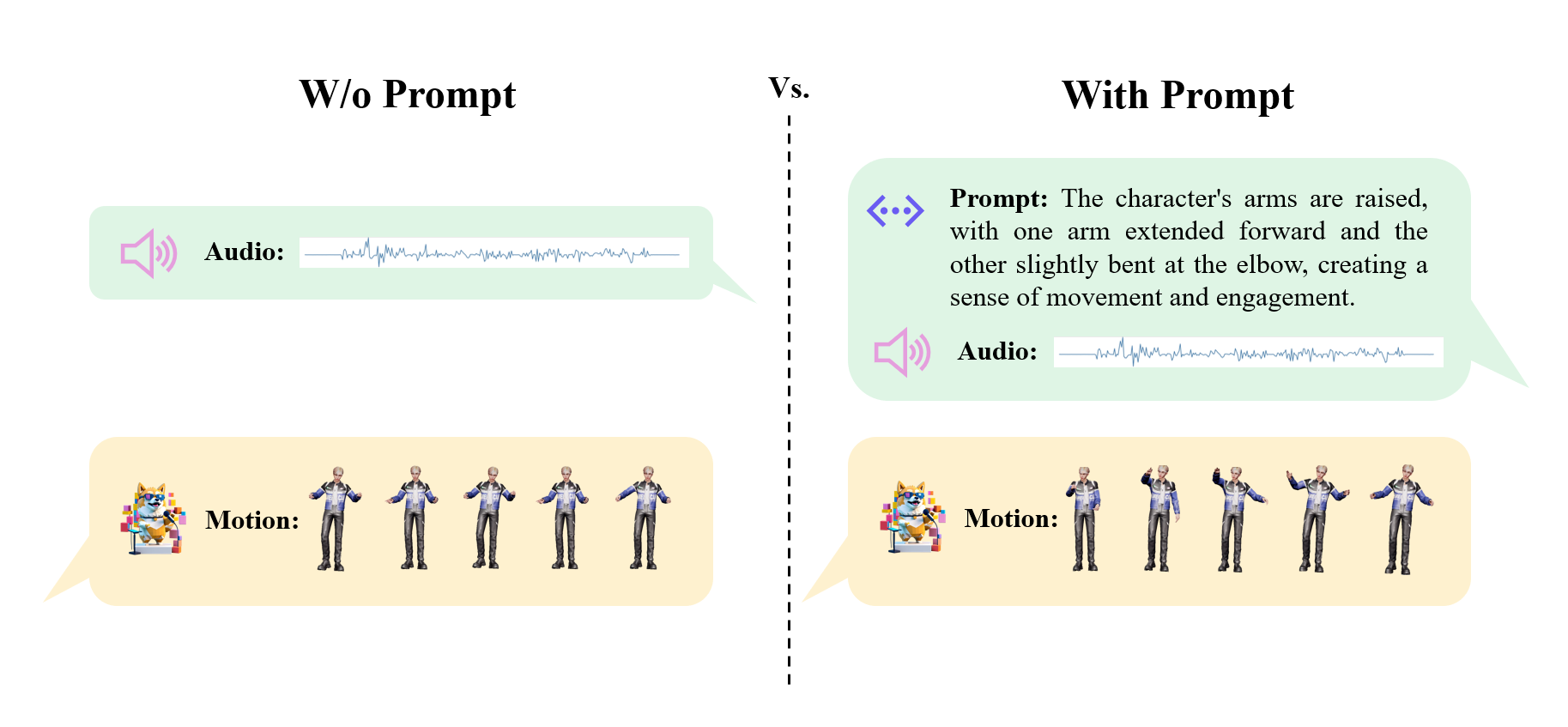}
    \caption{Comparison of motion generation with and without text prompts under the same speech input. The left side demonstrates the editing effect without text prompt, while the right side shows the results achieved by incorporating a text prompt.}
    \label{fig:results_compare}
\end{figure}
Subsequently, we continue fine-tuning on subtasks such as generating motion from audio and generating motion based on prompts and audio. Specifically, \textbf{the audio-driven motion generation} task is modeled as learning the conditional distribution $P(\hat{\boldsymbol{M}} \mid \hat{\boldsymbol{A}})$, aiming to generate a motion sequence conditioned on a given audio sequence.


For the task of \textbf{generating motion based on prompts and audio}, we model $P(\hat{\boldsymbol{M}} \mid \hat{\boldsymbol{A}}, \hat{\boldsymbol{T}})$, where the model generates motion sequences conditioned on both the audio sequence and textual prompts.

Our data composition is illustrated in Figure~\ref{fig:Sysoverview}. Since this task focuses primarily on motion generation, we perform supervised fine-tuning solely on the motion token portion. The fine-tuning objective is to minimize the negative log-likelihood loss:

\begin{equation}
    \mathcal{L}_{\text{SFT}} = - \sum_{t=1}^{T_M} \log P(\boldsymbol{m}_t \mid \hat{\boldsymbol{A}}, \hat{\boldsymbol{T}}, \boldsymbol{m}_{<t}),
\end{equation}

where $\boldsymbol{m}_{<t}$ denotes the sequence of motion tokens preceding time step $t$. By fine-tuning the pre-trained LLM in this manner, we leverage its powerful language modeling capabilities to generate coherent and contextually appropriate motion sequences conditioned on audio and textual inputs.




\section{Experiment}

\subsection{Dataset}
\textbf{Co-Speech Motion Dataset.} We utilize the BEAT dataset~\cite{liu2022beat} and the ZEGGS dataset~\cite{ghorbani2022zeroeggs} for training. The BEAT dataset is a high-quality corpus of speech motion data, featuring 30 characters and over 70 hours of motion data, along with corresponding audio, dialogue data, and annotations for key motion nodes. For the MotionRVQ, in alignment with the base-line setting in CaMN \cite{liu2022beat} , we select characters with speaker IDs 2, 4, 6, and 8 for training. For the LLM, we tokenize the motions of all characters using the aforementioned MotionRVQ and tokenize the audio from all characters using Encodec~\cite{defossez2022highfi}.  The ZEGGS dataset contains approximately 2 hours of motion data, encompassing 19 different styles. Due to the smaller size of the ZEGGS dataset, we utilize all available data for MotionRVQ training and LLM fine-tuning.

\begin{figure}
    \centering
    \includegraphics[width=0.9\linewidth]{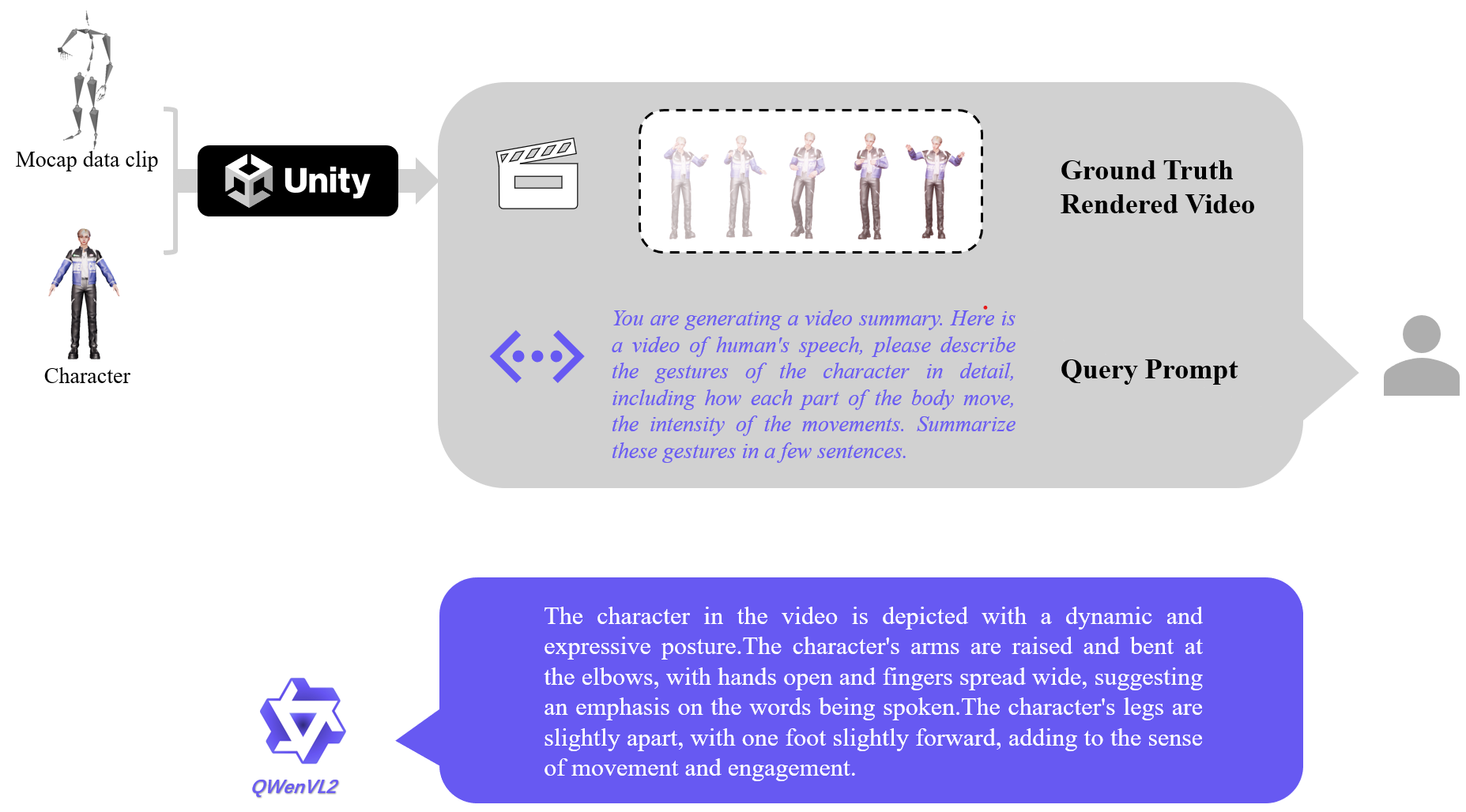}
    \caption{We bind the motion capture clips from the dataset to characters and use Unity to render them into videos, which are then input to the VLLM to generate textual descriptions of the motions.}
    \label{fig:vllm}
\end{figure}

\noindent \textbf{Generated text prompt dataset.}We retarget all motions from the BEAT dataset to our character assets and render them into video using Unity.  We use different characters for different gender of speakers in the dataset. Subsequently, we slice the videos and annotate each slice using the QWen VL2~\cite{qwen} 7B model to obtain natural language descriptions of the video content (Fig. \ref{fig:vllm}).

\subsection{System Setup}

%
%
Regarding the MotionRVQ, we employ a three-layer convolutional neural network and a six-layer Transformer architecture as the encoder, with a symmetric structure serving as the decoder. Our codebook size is set to 512, and the embedding space size is also 512. The depth of our residual layers is 4. We utilize a learning rate scheduler with a cosine decay and warmup, and employ the AdamW optimizer with a learning rate of \(2 \times 10^{-4}\) for training. Training the MotionRVQ on the ZEGGS dataset requires 2 A100 GPUs for 10 hours, whereas the same setup takes over a day to complete training on the BEAT dataset.

For the LLM model, we select the QWen1.5 series and conduct experiments at multiple scales, including QWen-1.5 0.5B, QWen-1.5 1.8B, QWen-1.5 4B, and QWen-1.5 7B. Each scale is trained on the data where the action mode is Speech, specifically using data from characters with IDs 2, 4, 6, and 8, along with the full dataset to pretrain. During the fine-tuning phase, we utilize a learning rate of \(2 \times 10^{-5}\) and conduct training for three epochs, with the longest training duration reaching 14 hours on 8 A800 GPUs. During inference, we adopt the technique proposed by Yao et al \cite{yao2024moconvq} as a post-processor to alleviate the foot-sliding problem.


In this section, we compare our framework with state-of-the-art systems to demonstrate the advances made in this work. We choose CaMN \cite{liu2022beat} and MultiContext \cite{Yoon_2020}, which are two recent successful 3D co-speech gesture synthesis frameworks, for comparison. We conduct quantitative evaluations using established metrics as a supplementary reference. Following the evaluation methods in previous works, 
we also conduct a perceptual user study to evaluate the overall quality of our synthesis results w.r.t. other methods.  The evaluation was conducted on the testset of BEAT dataset. Regarding methods trained on the ZEGGS dataset, the related codes for prior works are not available at the time of writing this paper. Consequently, we have included the rendered outputs of our model trained on the ZEGGS dataset in the supplementary videos.

\subsection{Quantitative Results}
We quantitatively evaluate the generated results with existing metrics including \textit{FGD}, \textit{Beat Alignment Score} \cite{li2021aichoreographermusicconditioned}, and \textit{Diversity}.  \textit{FGD} is calculated by comparing features of ground truth data and generated data, where the features are extracted by an auto-encoder network that is trained for reconstruction task. \textit{Beat Alignment Score} was introduced by \cite{li2021aichoreographermusicconditioned} to measure the alignment of audio beats and motion beats. \textit{Diversity} measures the average distance between each pair of gesture samples in the test dataset using the Euclidean distance metric. 

The test set comes from the BEAT dataset including 1183 5-second unseen motion sequences of 4 speakers. As shown in Table \ref{tab:evaluation_results}, 
our method achieves a better \textit{FGD}. The \textit{Diversity} and \textit{Beat Alignment Score} of ours are closer to the gound truth.

\begin{table}[ht]
\centering
\begin{tabular}{l p{2.5cm} p{2.5cm} p{2.5cm}}
\toprule
 & FGD $\downarrow$ & BeatAlign $\rightarrow$ & Diversity $\rightarrow$ \\ 
\midrule
GT & - & 0.93 & 14.30 \\ 
MultiContext & 176.2 & 0.98 & 26.71 \\ 
CaMN & 123.7 & 0.98 & 9.66 \\ 
QWen-7B (Ours w/o pretrain) & \textbf{119.4} & 0.91 & 7.42 \\ 
QWen-7B (Ours) & \textbf{47.0} & 0.92 & 11.55 \\ 
\bottomrule
\end{tabular}
\caption{Quantitative evaluation results of our method compared to existing approaches on the BEAT dataset. The metrics include FGD (calculated using our pre-trained auto-encoder), Beat Alignment Score, and Diversity, indicating the performance across different aspects of motion generation.}
\label{tab:evaluation_results} 
\end{table}

\subsection{User Study}

Following the method in previous works \cite{ConvoFusion24, Ao_2022}, we conduct user studies using pair-wise comparison. We create a dataset comprising 30 video pairs, with each pair containing our model's output, and another model's output or ground truth (GT) for the same speech audio segment. For each test, participants view two 10-second videos, and are asked to choose their preferred video based on two aspects: \textit{human likeness} and \textit{audio alignment}. The source of each video is invisible to the participants, and the pairs of videos, as well as the videos within each pair, are randomly shuffled.

\begin{figure}[h]
    \centering
    \begin{subfigure}{0.45\textwidth}
        \centering
        \includegraphics[width=\linewidth]{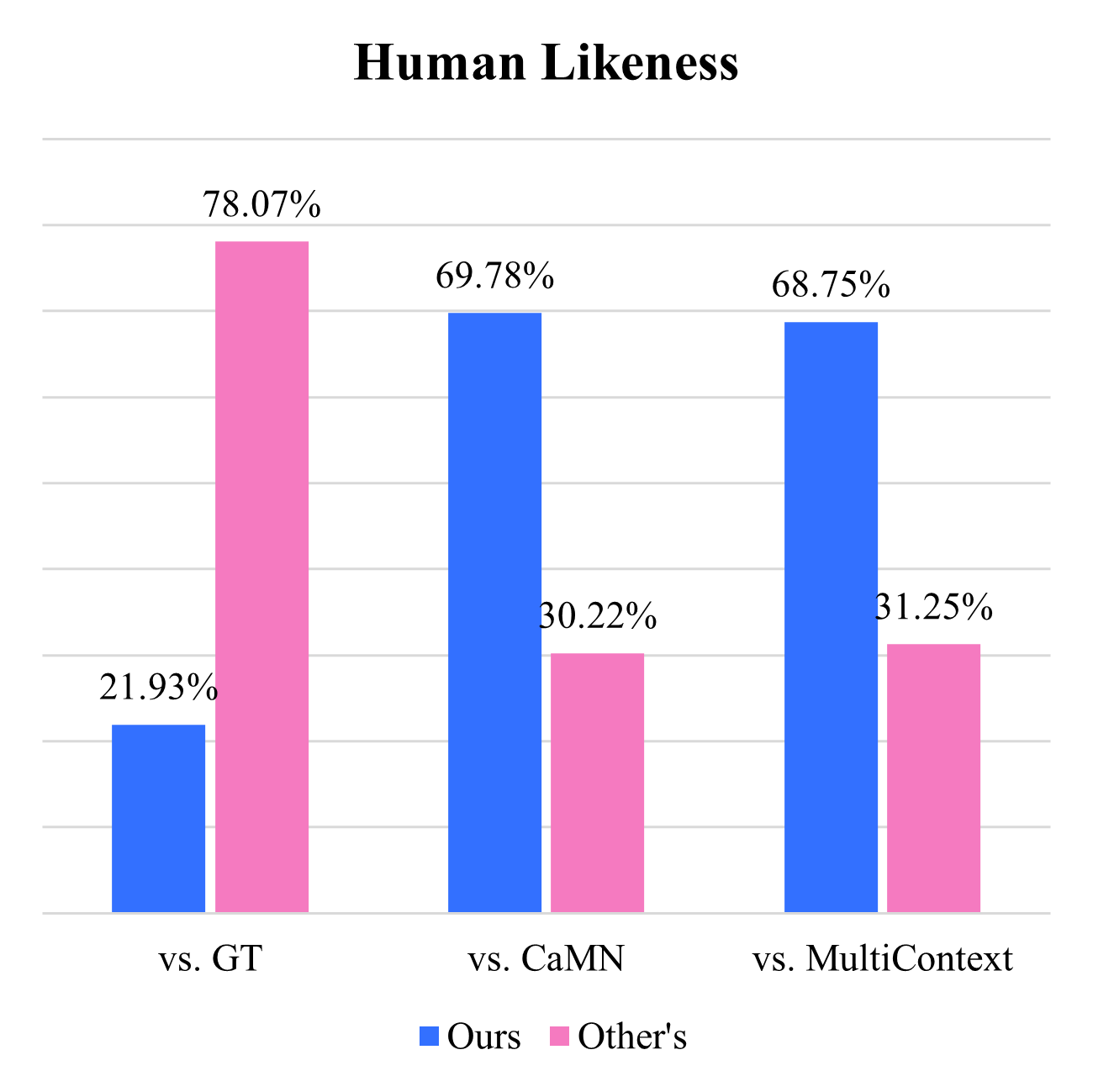} 
        \label{fig:HumanLikeness}
    \end{subfigure}
    \begin{subfigure}{0.45\textwidth}
        \centering
        \includegraphics[width=\linewidth]{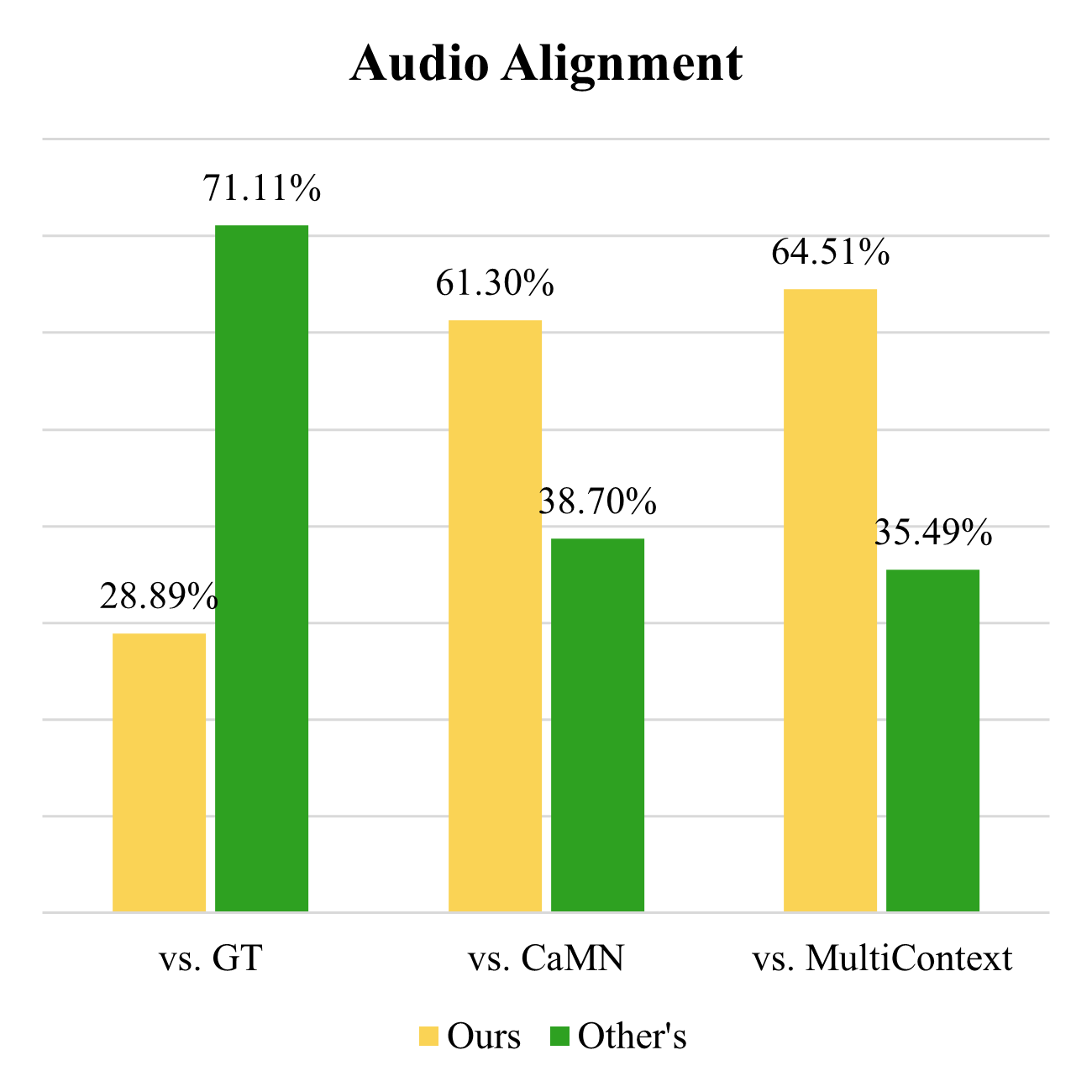}
        \label{fig:AudioAlignment}
    \end{subfigure}
\caption{Result of our user study. We compare our result with GT, CaMN and MultiContext respectively, on the aspect of human likeness and audio alignment.}
\label{fig:UserStudyResult}
\end{figure}

We recruit 13 volunteer participants. The result are plotted in figure \ref{fig:UserStudyResult}. As shown in the figure, our results are preferred for both human likeness and audio alignment, compared with CaMN and MultiContext. Meanwhile, there is still a distance between our result and GT.

\subsection{Ablation Study}

\begin{figure}[ht]
    \centering
    \includegraphics[width=0.8\linewidth]{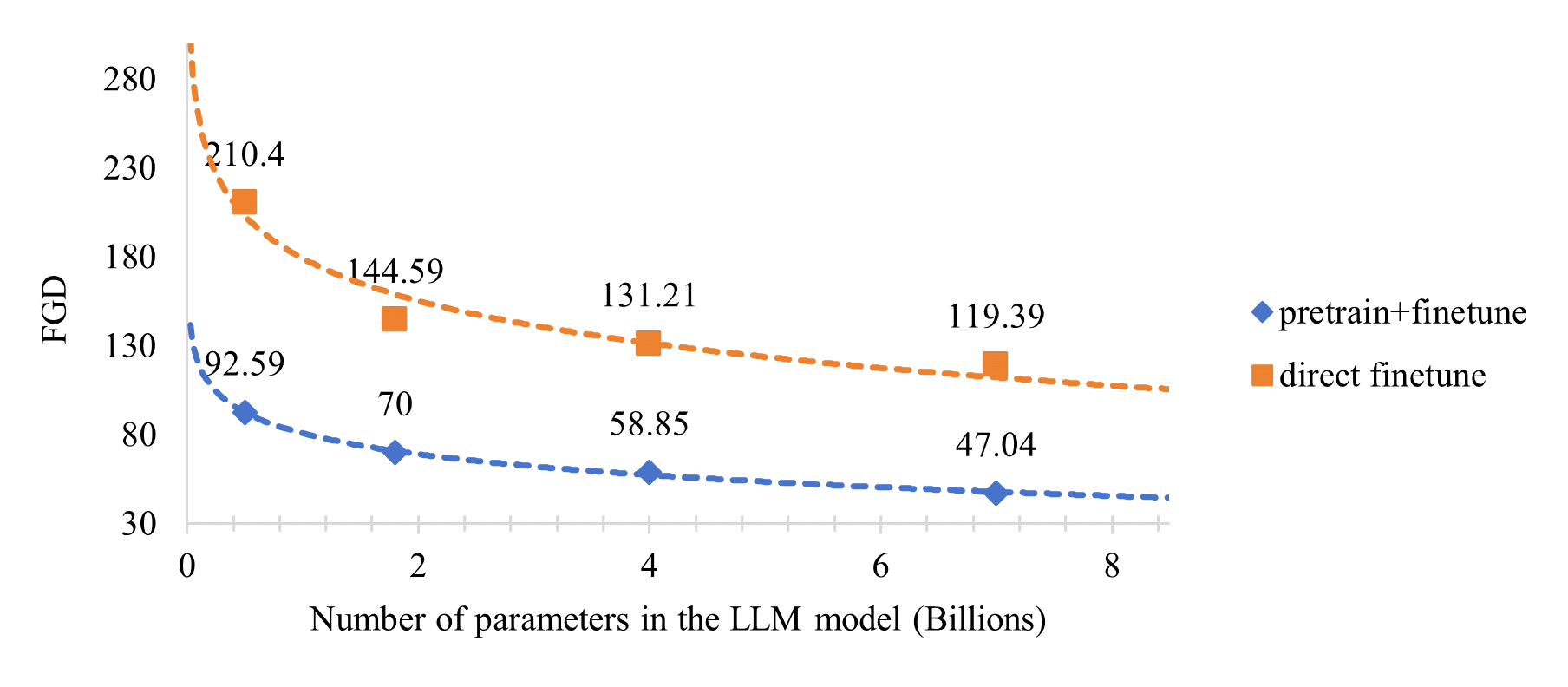}
    \caption{\textbf{This figure illustrates the impact of model size on the FGD scores of generated motions}. We conduct a series of experiments on the QWen1.5 series models with sizes of 0.5B, 1.8B, 4B, and 7B, utilizing fine-tuning configurations with datasets for speakers 2, 4, 6, and 8, as well as configurations involving pre-training followed by fine-tuning.}
   \label{fig:scale_comparison}
\end{figure}
\textbf{Validation of the scaling law.}
We trained LLM models of different scales under the same configuration, as shown in Figure \ref{fig:scale_comparison}. The results indicate that larger models produce higher quality action sequences. In contrast, smaller models often face issues such as sequence repetition and excessively long sequences. This demonstrates the importance of model scale in generating diverse and coherent action sequences.

\noindent\textbf{Training from scratch versus LLM Finetuning.}
The primary reason for choosing to construct the system using large language models (LLMs) is that training a model from scratch often fails to achieve a sufficient understanding of language. We conducted a simple validation experiment by constructing a motion sequence prediction network using 12 layers and 12 heads of Transformer architecture, with both text and audio as inputs. The experimental results indicate that the text has a minimal guiding effect on motion generation. This issue primarily arises from the need for more data to adequately understand the text. Therefore, we opted to use pre-trained LLMs, which have been extensively trained on large text datasets and possess the capability to understand text sequences.

\section{Conclusion}

In this paper, we present \textit{LLM Gesticulator}, a  novel framework to synthesize controllable co-speech gestures. With the goal of generating natural, rhythmic, and controllable gestures, we formulate the co-speech generation task as a sequence-to-sequence translation problem so that by proper tokenization of data in different modalities, we could leverage pre-trained large language models to accomplish this task. We conduct experiments on LLM models of different scales and observe performance gains  as the model size increases. This finding is consistent with the scaling law and demonstrates the scalability of our approach. We render and annotate the motion clips in BEAT dataset and use the video content description as text prompt in our training framework,  thanks to the text understanding ability of pre-trained LLM models, our method demonstrates strong text controllability where the style and content of generated gestures can be controlled by user prompts. Our method outperforms prior works on existing quantitative evaluation metrics and is more preferable in user studies in terms of \textit{human likeness} and \textit{audio alignment}.

However, there is still room for improvement in our framework. Our method cannot achieve real-time stream inference. We believe that with the advancement of acceleration techniques such as quantization and distillation,  and better engineering practices, real-time stream co-speech synthesis is possible to achieve. In addition, we consider the text and audio modalities of input data in this work, but our framework can be easily extended to support other modalities such as images or videos by proper tokenization and alignment process. Generating full-body gestures that include facial-expressions is also an interesting topic, and we left them as directions worth exploring in the future.

\newpage
\appendix    

\section{More Visualization Results}
\begin{figure}[h]
    \centering
    \includegraphics[width=1\linewidth]{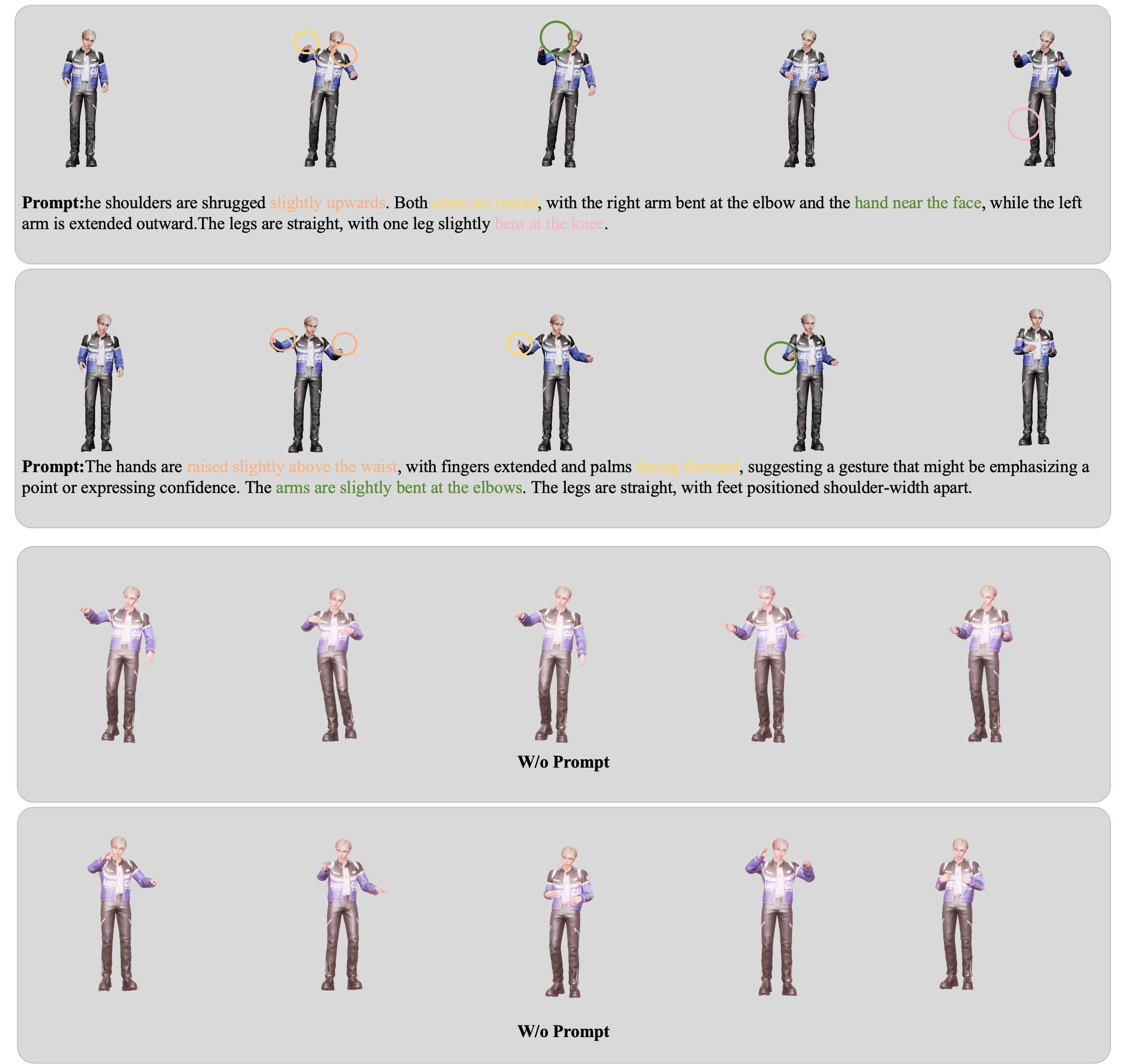}
    \caption{Co-speech motion generation with prompt and without prompt}
    \label{fig:res1}
\end{figure}
\begin{figure}[H]
    \centering
    \begin{subfigure}{0.4\textwidth}
        \centering
        \includegraphics[width=\linewidth]{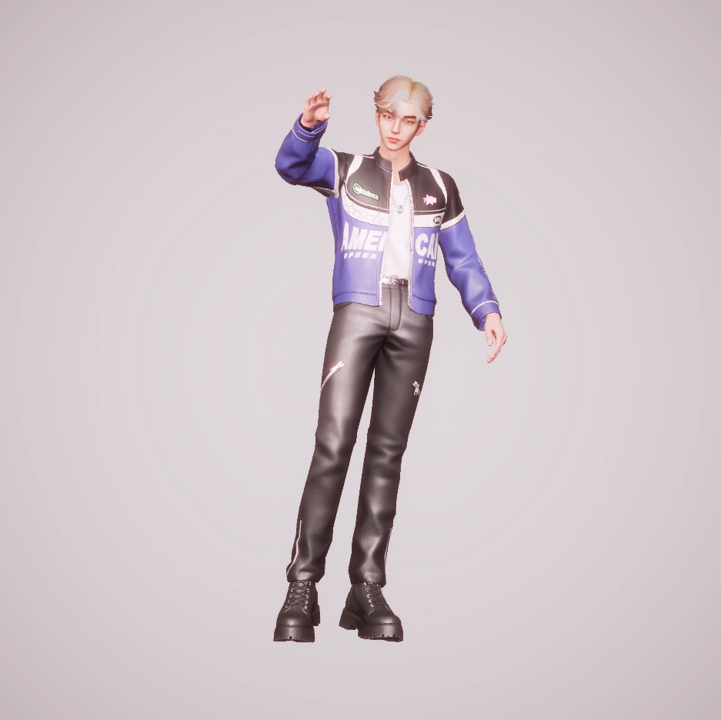} 
        \label{fig:RenderPreview_Taeho}
    \end{subfigure}
    \begin{subfigure}{0.4\textwidth}
        \centering
        \includegraphics[width=\linewidth]{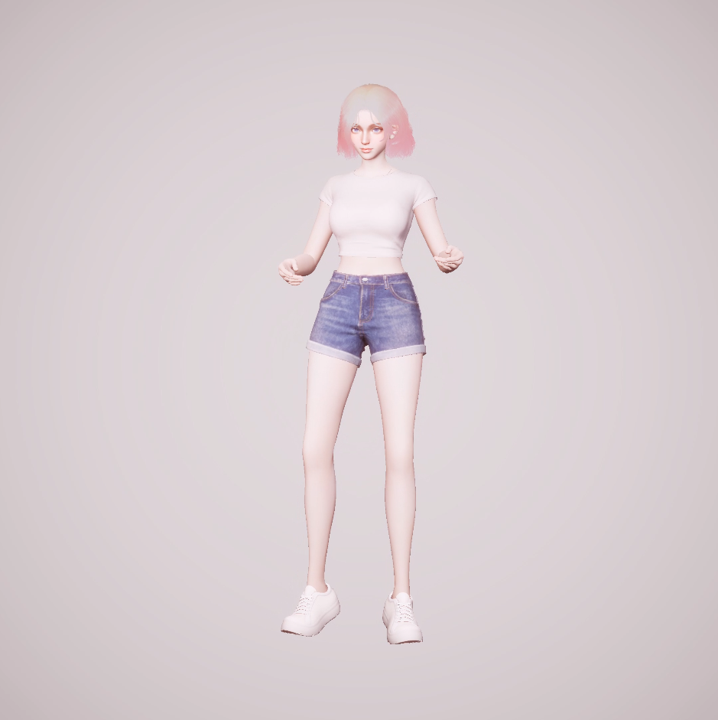}
        \label{fig:RenderPreview_Rena}
    \end{subfigure}
\caption{Preview of our rendered video for VLLM labelling. Different characters are chosen for different gender of speakers in the dataset.}
\label{fig:VideoRenderPreview}
\end{figure}

 

\bibliography{report} 
\bibliographystyle{spiebib} 

\end{document}